# Systems engineering applied to ELT instrumentation: The GMACS case

D. M. Faes*[a], A. Souza[b], C. Froning[c], L. Schmidt[d], D. Bortoletto[a,e], E. Cook[d], D. L. DePoy[d], T.-G. Ji[f], D. Jones[g], H.-I. Lee[f], J. L Marshall[d], C. M. Oliveira[a], S. Pak[f], C. Papovich[d], T. Prochaska[d], R. A. Ribeiro[a], K. Taylor[h]

[a]Departamento de Astronomia, IAG, Universidade de São Paulo - São Paulo, Brazil;
[b]Universidade Federal de Minas Gerais- Belo Horizonte, Brazil;
[c]McDonald Observatory, University of Texas at Austin - USA;
[d]Department of Physics and Astronomy, Texas A&M University - USA;
[e]Instituto Mauá de Tecnologia - Mauá, Brazil;
[f]School of Space Research, Kyung Hee Univeristy - South Korea;
[g]Prime Optics - Australia;
[h]Instruments4 - California, USA.

**ABSTRACT**

An important tool for the development of the next generation of extremely large telescopes (ELTs) is a robust Systems Engineering (SE) methodology. GMACS is a first-generation multi-object spectrograph that will work at visible wavelengths on the Giant Magellan Telescope (GMT). In this paper, we discuss the application of SE to the design of next-generation instruments for ground-based astronomy and present the ongoing development of SE products for the GMACS spectrograph, currently in its Conceptual Design phase. SE provides the means to assist in the management of complex projects, and in the case of GMACS, to ensure its operational success, maximizing the scientific potential of GMT.

## 1. INTRODUCTION

The development of innovative scientific instrumentation has a number of challenges, during the stages of design, construction, and long-term operation. Astronomy is no exception. An important tool for the development of the next generation of extremely large telescopes (ELTs) is applying rigorous Systems Engineering (SE) process and practices.

Multi-Object Spectroscopy (MOS) is an efficient, highly productive observational mode for astronomical research. Coupling a MOS with the next generation of ELTs will provide new windows for scientific discoveries. Good summaries of the science cases for MOS using the ELTs can be found in Colless (2006), Neichel et al. (2006), and Evans et al. (2015).

In this context, GMT is developing the Giant Magellan Telescope Multi-object Astronomical and Cosmological Spectrograph (GMACS). GMACS is a multi-object spectrograph working at visible wavelengths for the GMT. See DePoy et al. (2018) for a project status overview.

In Section 2 we present a background on Systems Engineering and its importance for projects such as the ELTs. Section 3 has a brief discussion of the challenges involved in the development of instrumentation for the ELTs, with emphasis on spectrographs. The Systems Engineering processes of the GMACS Conceptual Design phase are described in Section 4. The next phases for GMACS and SE challenges are briefly described in Section 5. Our final remarks are in Section 6.

*Further author information: (Send correspondence to DMF: Departamento de Astronomia IAG/USP -- Rua do Matão, 1226, Cidade Universitária - São Paulo/SP, Brasil - 05508-090). E-mail: moser@usp.br, Telephone: +55 11 3091 2705

## 2. SYSTEMS ENGINEERING

### 2.1 Goal

Systems Engineering (SE) proposes a series of methodologies and practices to ensure the successful development and operation of systems. Historically, many of the SE process applications were in the aerospace industry and the defense industry (INCOSE 2015). However, modern SE has a broader scope of applications (e.g., Product-SE, Enterprise-SE, Service-SE, etc). For a discussion of the impact of SE in ground-based observatories, see Swart and Meiring (2003).

Some of the reasons that led to the implementation of SE methodology in complex projects are: (i) Limited product effectiveness; (ii) Results often unrelated to the actual needs; (iii) Serious delays in schedules; (iv) Excessive costs; (v) Bad development directions; and (vi) Need for unification or standardization of practices created in different fields.

The early implementation of SE practices aims to guarantee a good understanding of the needs and requirements of the system from concept to disposal. SE design methodology will widely consider the system life cycle, the needs of the final users, and mitigate risks as early as possible by working closely with specialized engineers for design decisions.

Figure 1 illustrates the importance of the application of SE throughout cost allocation per phase and expenses in case of defect. This pattern has been observed in different projects from different domains and has been used to justify the intense use of SE practices at the concept and design phases of a project, since those phases are where decisions will allocate most of the cost of the project and errors will be corrected with less expense.

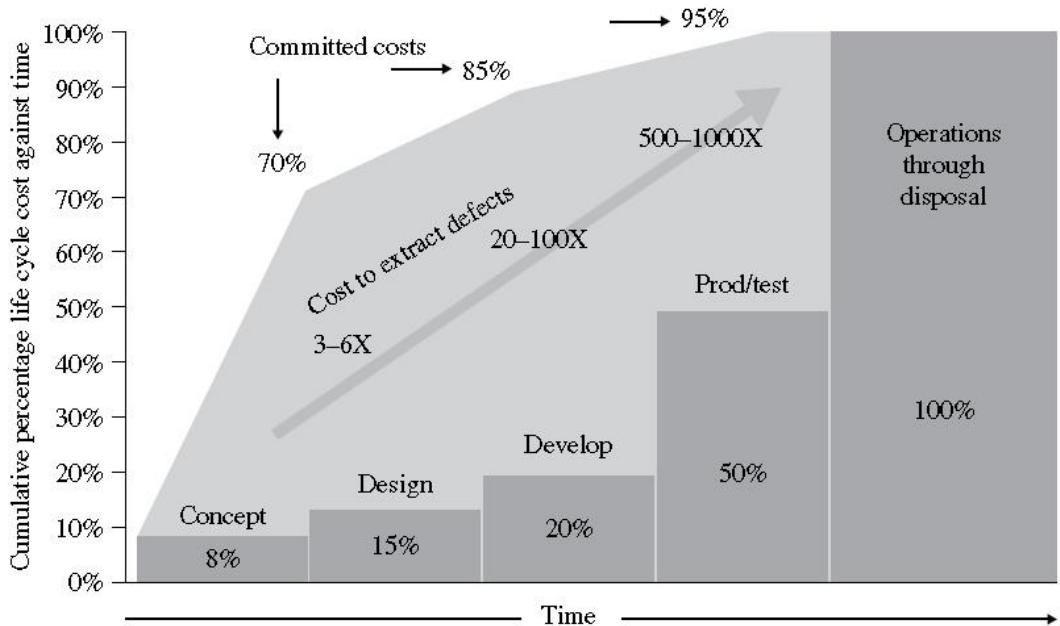

Figure 1. Costs evolution through project life cycle from INCOSE Handbook 4th Edition.

# 3. SPECTROSCOPY ON THE ELTS

The increasing cost of ground-based astronomy with the size of the telescope apertures, coupled with increased technical complexity, are important reasons for observatories to seek the support of SE practices. Table 1 contains data from the survey of van Belle et al. (2004), as well as GMT as a representative case of the cost of an ELT.

Table 1. Comparison of telescope projects. Costs are approximated and based on the buying power of the US dollar in 2000.

| Telescope | Diameter | Project Start | First Light | Cost |
|---|---|---|---|---|
| Magellan (2x) | 6.5 m | 1994 | 2002 | $135.4M |
| Keck II | 10 m | 1991 | 1996 | $85.9M |
| Keck I | 10 m | 1985 | 1993 | $139.1M |
| MMT | 6.5 m | 1979 | 2000 | $49.4M |
| GMT | 24.5 m | 2006 | 2024[1] | TBD[2] |

The driving goal of constructing spectrographs for ELTs it to enable spectroscopy of targets that are currently only visible through images, like primordial (high-redshift) galaxies. ELTs are also excellent tools for obtaining high cadence of observations of transient events, such as transit of exoplanets. However, the construction of instrumentation for these and other MOS instruments has a number of challenges. Here we list the needs of ELT spectrographs that should be addressed in a systemic way.

## 3.1 Scale up to keep FoV

One of the main difficulties in the construction of spectrographs for ELTs, particularly in the seeing-limited regime, is the physical size of the optics. By construction, the working f-number of reflector telescopes do not change considerably with its size. This means that the physical size of the generated images grows linearly with the diameter of the telescope. Table 2 contains typical values for telescopes with f-number ≈16. For ELTs covering a reasonable wide field of view, one can expect images that are more than one meter in size!

Table 2. Comparison of the plate scale and a 10 arcmin (') image of different telescopes aperture sizes and the usual effective f-number ≈ 16 (approx. values). The GMT values, with effective f-number = 8.16, are indicated with the ∗ sign.

| Diameter | Plate Scale | Size of 10' |
|---|---|---|
| 3 m | 4.0 "/mm | 150 mm |
| 10 m | 1.3 "/mm | 450 mm |
| 24.5 m* | 1.0 "/mm* | 600 mm* |
| 30 m | 0.4 "/mm | 1350 mm |

## 3.2 Competitive resolution and spectral coverage

When the resolution of the generated spectra is considered, there is a similar impact. The main factor controlling the spectral resolution in terms of the size of the optics is the ratio between the diameters of the collimator and the telescope (Allington-Smith 2007). Because it is very difficult to create large lenses, in first order the resolution of a given spectrograph is inversely proportional to the diameter of the telescope.

---

[1] Representative predictive schedule

[2] See Fanson et al. (2018) for a project status overview.

### 3.3 High mechanical stability

The size of the optics generates large instruments. Spectrographs, particularly those located at the gravity-variant focus position will need real-time mechanical actuators to correct mechanical flexure with gravity vector changes. This is true for GMACS, which will be installed below the telescope mount, but also for spectrographs in Nasmyth focus that need to rotate accordingly to the observed field. The total mass of the instruments increases the chances of inaccurate flexure corrections that can greatly degrade the efficiency and quality of the generated spectra.

### 3.4 Integration with AO capabilities

The integration with adaptive optics resources simultaneously serves to identify and observe faint targets as well as provides an effective mechanism to increase the resolution of the generated spectra. The area of the primary ELT mirrors generates additional deflections for adaptive optics corrections, especially if it encompasses multiple targets or a large field of view.

### 3.5 High throughput

High throughput is a challenge in a large system (which often uses internal reflections to reduce its volume) and it is still integrated into an adaptive optics system. This is important for the telescope to fulfill its purpose of observing fainter targets, since the goal is to combine instrumental efficiency with the telescope's collecting area (a high throughput instrument on a 10 m telescope can be equivalent to a low throughput instrument in a 30 m telescope).

## 4. GMACS AS A SUBSYSTEM OF THE GMT

As mentioned in Section 2, SE methodology aims to address any issues of the project as early as possible. We describe here this methodology in more detail, focusing on describing its tailored version as applied to GMACS.

The GMT System Engineering Management Plan defines the project hierarchy, overall scope of each project phase and highlights the common artifacts recommended to be used when implementing requirements flow-down, interface definition, risk analysis, planning, decision analysis and cost estimates (Angeli et al., 2018).

GMT recommends this approach to all instrumentation groups. Similar to GMACS, a novel systems engineering approach is being applied to the GMT-CfA Large Earth Finder (GCLEF) Podgorski et al. (2014).

### 4.1 Top-down approach

The Top-down approach covers managerial and design practices. It is a way of managing and designing the project so that engineers can address first architectural aspects of the project without focus on detail. As more information becomes available, details will be addressed in the design. To start this SE seeks to capture all subsystems necessary; for that, a PBS (Product Breakdown Structured) is developed together with engineers. The PBS will help manage the group, plan activities, and organize the flow-down of requirements from system to subsystem.

GMACS' overall functional architecture, shown in Figure 2, describes the main subsystems addressed in the conceptual design to help capture the key technical functions for GMACS, including its interface with MANIFEST (Many Instrument Fiber System) and long term functions, such as AITC (Assembly, Integration, Test and Commissioning).

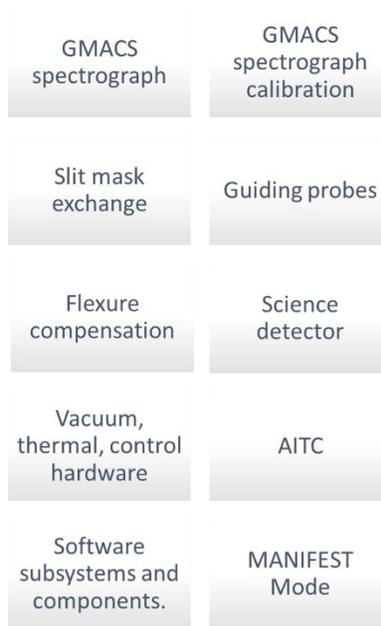

Figure 2. GMACS Functional Architectural Decomposition.

## 4.2 Traceability of requirements and requirements flow-down

The requirements flow-down for GMACS is the responsibility of the systems engineer with the support of specialized engineers and astronomers. It starts from the identification of scientific cases, operational aspects and constraints imposed by the observatory. From these, the first flow-down is written and the initial requirements that will guide the technical team captured. This process is illustrated in Figure 3.

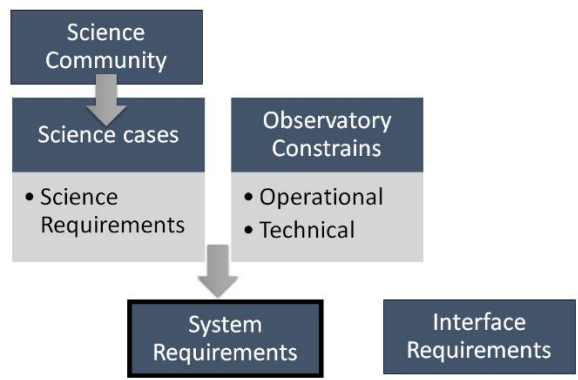

Figure 3. GMACS Top Down approach for requirements flow-down.

To date, we have captured almost a hundred requirements, classified as architecture, interface, and performance types, illustrated in Figure 4. For the architecture type we have captured requirements that are constraints, functions or quality aspects. For the interface type we have requirements that concern the interface with the observatory, telescope, and other instruments. For the performance type we have requirements that concern instrument performance and can be directly linked to design solutions.

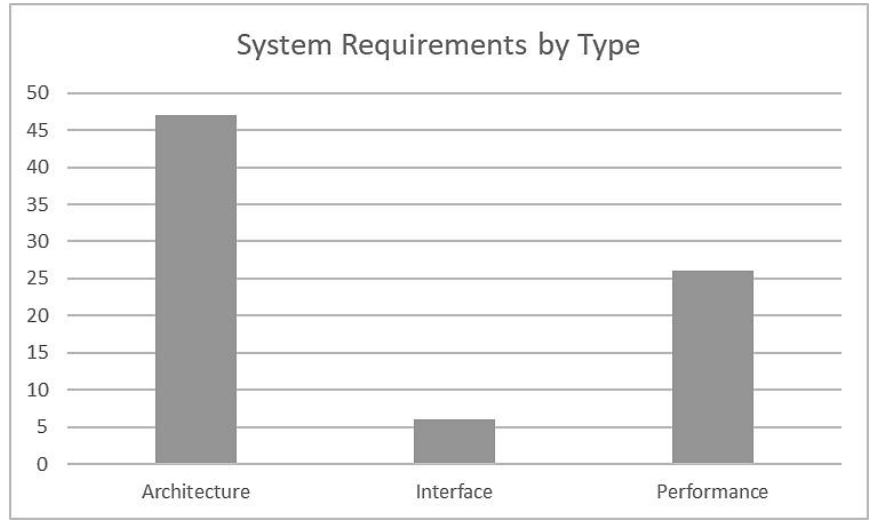

Figure 4. System requirements status.

An overview of the current GMACS system requirements is shown in Figure 5 and illustrates those requirements considering the entire GMACS life cycle as expected from conceptual design. Regarding traceability, system requirements will be linked to subsystems requirements that are being generated by the GMACS engineers as the design decisions advance.

The Traceability from all identified aspects and the derived requirements are managed by the systems engineer, using requirements management tools, such as ReqView® and DOORS® and resource estimation tools such as WBSpro®. The traceability of the requirements allows the estimation of impact of changes, to know better the scope of the project, to justify decisions, besides helping estimate cost and schedule.

### 4.3 Record of decisions, knowledge management

From the initial flown down of requirements many design concepts are possible and the experience of the group and research are important to make decisions that will reduce the possibilities and narrow down those to the most likely options. The systems engineers at GMACS oversees those decisions and participate in group meetings to make sure the complete life-cycle is considered, instead of only performance and cost. All decisions are documented as trade-off or technical notes from templates developed by the SE team.

### 4.4 Cost and schedule estimates

Systems engineering applied to cost and schedules estimates will consider the entire system's life cycle (development, fabrication, integration, validation, commissioning, operation, upgrade, and disposal), in addition to social and environmental aspects that may influence in some stage of the system life-cycle.

Making long term estimates, is a matter addressed statistically by SE tools. In the GMACS case, to deal with the possibilities and uncertainty of the six design and development remaining phases until first light, the risk analysis will be used as input for the estimation tools.

### 4.5 Interfaces

Interface is one of the most challenging aspects that SE deals with. It requires communication, organization, discipline and broad understanding of the overall aspects of the system, its subsystems, operation and environment, as illustrated in the context diagram in Fig. 6. It takes into account, for example, mechanical interfaces such as the GMT DG Mount, which is the place of installation at the Direct Gregorian beam of the telescope and the means of installation, which in the GMACS case is through the telescope Pier lift.

| Requirement Title | MANIFEST Interface Function | System Architecture | GMACS spectrograph function | Slit mask exchange function | Guiding probes function | Flexure Compensation subsystem | Mechanical interfaces function |
|---|---|---|---|---|---|---|---|
| MANIFEST integration | Control hardware function | Operations software function | Acquisition software Function | Data calibration function | AITC and maintenance Function | Environmental isolator function | Science detector function |
| Tracking Error | Focus Error | Mounting Piston (Focus) Error | Mass Limit stowed | Scheduled Time off | Guiding probe Pointing Accuracy | N2 Source | Liquid Nitrogen Source |
| Observatory Interface | Availability | Safety | Safeguards | Electronic Cabinets | Night Maintenance | AITC and Maintenance | Utility Interface |
| Cable Trays | Mechanics Interfaces | Instrument Availability | Proposal Preparation Tools (Operations SW) | Automated Startup/Shutdown | Science Data Format (Acquisition SW) | Data Pipeline (Data Calibration SW) | Data Quality Assessment (Operations SW) |
| Telemetry (Control/Operations SW) | Health Monitoring (Control/Operations SW) | Instrument Vibration Budget | Installation | Equipment Service Access | Lifetime | DRAFT User Interface | Engineering Mode |
| Sampling Rates | Diagnostic Software | Diagnostic Telemetry Time Stamp | DRAFT On-Axis Natural Seeing Image Size | DRAFT Science Target Acquisition Uncertainty | Spectrograph Low resolution | Spectrograph High resolution | Spectrograph main optics |
| Spectrograph Channels | Channel Transition | Spectrograph Throughput | Long Slit Mode | Detector Constrains | GMACS spectrograph function | Red Detector Assembly | Optical Transparency Budget |
| Detector Quantum Efficiency | Detector performance | Power modes | Sky Coverage | Astrometric Variation | Observation time constrain | Wavelength lower limit | Observation performance image |
| Observation performance detection | Observation Sensitivity Variation | Observation flux in-field | Observation flux relative | Tracking targets | Time accuracy | Image Quality Variation | Spectroscopic stability |

Figure 5. GMACS System Requirements.

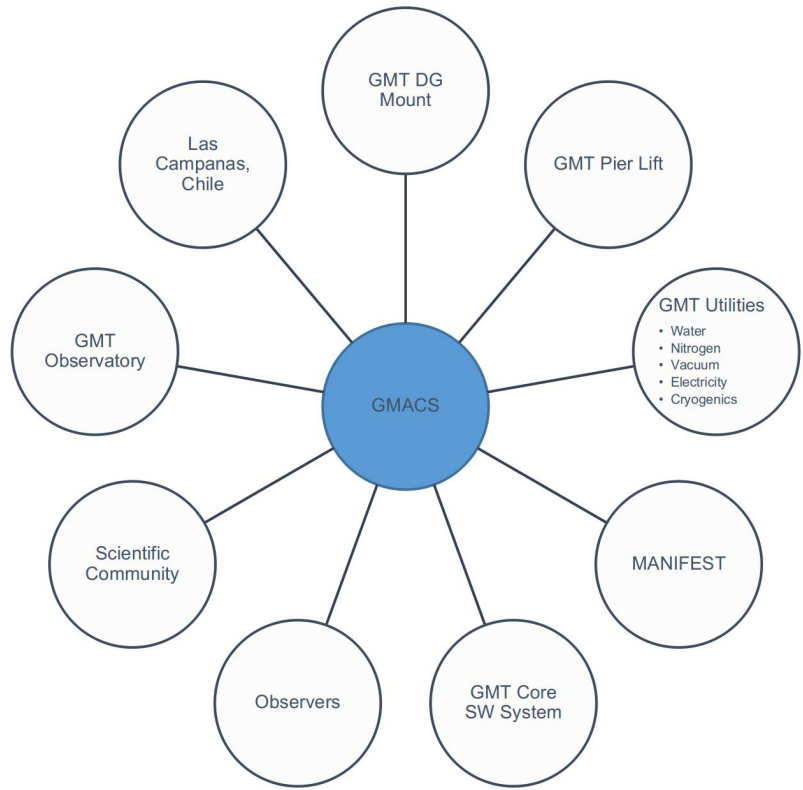

Figure 6. GMACS Context Diagram.

For a conceptual design like the one currently being undertaken for GMACS, top-down and bottom-up approaches need to be combined to consider all interfaces. Top-down allows the identification of interfaces from a wide point of view, considering observatory aspects, such as operation, facility instruments, and AITC. Bottom-up complements this by allowing the identification of interfaces that depend on subsystems solutions. In order to coordinate both approaches, good practices of requirements traceability and knowledge management need to be followed, which includes good communication between all stakeholders that SE needs to be prepared to facilitate.

### 4.6 Risk management

Risk Management of the project allows the project manager to better allocate resources for the project, but it is also supported by System Engineers. Their interest is to identify technical and strategic risks and to help the project manager plan mitigation activities that can be applied in early stages. When applied at conceptual design, such as GMACS, the awareness of the risks allows to mitigate most of them during the trade-off and decision process. For GMACS, the expectation at the end of the conceptual design is to have all risks from the red area (Figure 7), moved to yellow and green, meaning that the risk will be much more manageable.

![Risk matrix showing Impact (1-5) vs Likelihood (1-5) with colored cells. Red cells contain risk IDs: 03 at (5,4), 11 at (5,5), 01 at (4,4), 02 at (4,5), 07 at (4,5), 06 at (3,4), 10 at (3,4), 09 at (3,4), 04 at (3,5), 05 at (3,5). Yellow cell: 08 at (3,3).]

Figure 7. Example of risk matrix to identify and visualize the evolution of the risks. The higher the likelihood, more probable is its occurrence; the higher the impact, the greater the (negative) consequence in the project.

GMACS uses the same approach for risk management as GMT, only adapted for scaled scope, at metrics for cost and schedule impacts and likelihood. Following that approach, all risks are classified as technical, cost, and schedule and have the impacted requirement traced to it.

## 5. GMACS NEXT STAGES

GMACS is finalizing its conceptual design and achieving its deliveries. Highlighting the SE process applied and described in Section 4, we had analyzed all GMT input documentation, such as System Engineering Management Plan (SEMP), Operational Concept (OpsCon) and others, and we are going to provide for GMTO evaluation all the outputs illustrated in Figure 8.

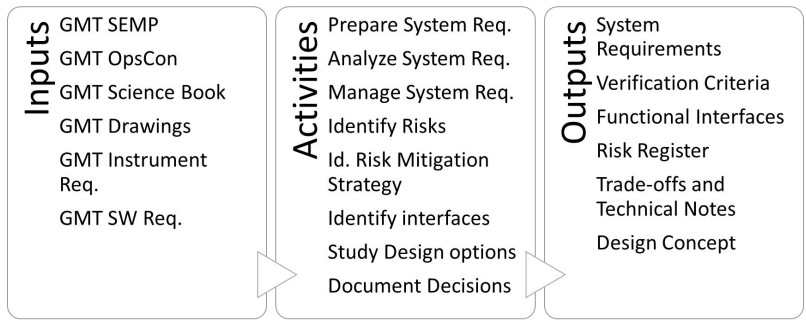

Figure 8. Conceptual design SE process

After approval, GMACS has a design process ahead that will be supported by SE process and practices. At Preliminary Design, robust risk management will continuum, system requirements will be refined, interface descriptions will be detailed, analysis will be done to understand quality and hazard aspects, and verification and validation plans will be defined. At Critical Design, operational aspects will be detailed and finalized regarding process, transportation and integration. At Manufacturing Readiness, quality assurance and safety plans are stablished. At Test Readiness, Pre-Shipment and Site Acceptance a system engineers overseas the process, ready to mitigate issues if necessary.

## 6. CONCLUSION

This work addresses the objectives of SE for complex systems and how SE is being applied to the GMACS Conceptual Design in order to mitigate the risks for the next phases. This is contextualized within SE processes recommended by GMT, and the focus is on the challenges of the multi-object spectroscopy technique for the ELTs need to overcome. From a broader perspective, it is pointed out how SE methods can assist the development of complex projects and maximize the scientific potential of big experiments, such as the ELTs.

## 7. ACKNOWLEDGEMENTS.

DMF acknowledges support from FAPESP grants 2011/51680-6 and 2016/16844-1.